\documentclass[aps,pre,amssymb,showpacs]{revtex4}

\usepackage{graphicx}
\usepackage{psfig}

\newcommand{\thf}{\theta_1(n+1)}
\newcommand{\ths}{\theta_2(n+1)}
\newcommand{\wexp}{e^{-i\omega}}
\newcommand{\wdw}{W^{\dagger}WU}

\begin{document}
\def\/{\over}
\def\<{\langle}
\def\>{\rangle}
\def\({\left(}
\def\){\right)}
\def\[{\left[}
\def\]{\right]}
\def\I{{\cal I}}
\def\i{{\rm i}}
\def\d{{\rm d}}
\def\e{{\rm e}}

\title{
Signatures of Classical Diffusion in Quantum Fluctuations of 2D Chaotic Systems}
\author{Tsampikos Kottos, Alexander Ossipov and Theo Geisel}
\address{{Max-Planck-Institut f\"ur Str\"omungsforschung und Fakult\"at Physik
der Universit\"at G\"ottingen,}\\
{Bunsenstra\ss e 10, D-37073 G\"ottingen, Germany}}

\date{\today }

\begin{abstract}

We consider a two-dimensional ($2D$) generalization of the standard kicked-rotor (KR) and 
show that it is an excellent model for the study of universal features of $2D$ quantum systems 
with underlying diffusive classical dynamics. First we analyze the distribution of wavefunction 
intensities and compare them with the predictions derived in the framework of diffusive  
{\it disordered} samples. Next, we turn the closed system into an open one by constructing a 
scattering matrix. The distribution of the resonance widths ${\cal P}(\Gamma)$ and Wigner delay 
times ${\cal P} (\tau_W)$ are investigated. The forms of these distributions are obtained for 
different symmetry classes and the traces of classical diffusive dynamics are identified. Our 
theoretical arguments are supported by extensive numerical calculations.
\end{abstract}
\pacs{05.45.Mt, 73.23.-b}
\maketitle

\section{Introduction}
\label{intro}

Random Matrix Theory (RMT) was invented more than 50 years ago by Wigner in order to 
describe the statistical properties of complex nuclei \cite{W55}. In recent years the fast 
developments in mesoscopic physics and quantum chaos gave a new boost to the RMT approach 
\cite{E97,S99}. The strength of RMT consists in  the universality of its predictions 
containing no energy or length scales or any parameter dependence. At the same time this 
is its weakness, because it does not allow to take into account various phenomena appearing 
in mesoscopic systems which introduce new length scales or parameters in the system.

The breakdown of {\it universality} and the traces of the underlying classical dynamics 
are important topics in quantum chaos studies and have major applications in mesoscopic
physics. It is therefore highly important to compare the results from RMT models, to the 
results from systems that have a 
semiclassical limit, and to look for circumstances where semiclassical tools rather than 
RMT become relevant. In this paper we extend our previous studies \cite{OKG02,OKG03} on 
two-dimensional chaotic systems with underlying classical diffusion paying attention to 
applications of our results for random lasers and microwave absorption. We identify the 
signatures of classical diffusion in quantum fluctuations by analyzing the distribution 
of wavefunction intensities of closed systems, and the resonance widths and Wigner delay 
times of their corresponding scattering analogs. 

Our analysis is based on a two-dimensional ($2D$) generalization of the well known Kicked 
Rotor (KR) model . For large values of the kicking strength the model is classically chaotic 
and the dynamics is diffusive. The detailed analysis of the classical system allows us 
to extract the relevant parameters that penetrate into the quantum mechanical description. 
The quantum analog shows {\it dynamical} localization, \cite{I90,FGP82,DF88,J03} with 
eigenstates which are exponentially localized in a similar fashion as the eigenstates of 
disordered systems. However here the localization occurs due to complicated interference 
effects created by the underlying classical chaotic dynamics. Attempts were made to put 
this analogy between the kicked rotor and  disordered systems \cite{FGP82, AZ96} on a solid 
base but the problem is  still not solved completely \cite{J03}. 

Since the pioneering work of Anderson \cite{And58} it is known that the diffusion in 
{\it disordered} systems can be affected very strongly by quantum localization. If the 
system size exceeds  the localization length, then the diffusion is completely suppressed 
after the time needed for the  wave packet to spread over the scale of the localization 
length. But even in the absence of strong localization the existence of {\it pre-localized} 
states  \cite{AKL91,MK95,FM94,FM95,FE95,FE95-2,M97,M00-2,SA97} influences the behavior 
of different physical quantities in the diffusive regime. Having this in mind, it is 
legitimate to ask, what happens with deterministic {\it chaotic} systems when quantum 
mechanical description becomes relevant? Our analysis shows that pre-localized states 
are present also in the case of dynamical systems and affect various statistical quantities 
like resonances and Wigner delay times. For the analysis of the latter quantities, we 
turned the closed KR to a scattering system by imposing `absorption' to a set of momentum 
states. The resulting scattering matrix is written in terms of the unitary evolution 
operator of the corresponding closed system and a projection operator. A detailed 
description of this construction is presented.

The paper is structured in the following way. In section II the classical model is  
introduced and the main definitions are given. Sections III and IV are devoted to the 
corresponding quantum mechanical system. The quantization on a torus is presented in 
section III, while in IV we present a detailed construction of the scattering matrix. 
The next two sections V and VI are dedicated to the analysis of various statistical properties
 of the closed and the open  $2D$ KR model correspondingly. 
Finally, our conclusions are summarized in section VII.

\section{The classical $2D$ Kicked Rotor}

A prominent example of a system with classical diffusion is, the well known in the field 
of quantum chaos, Kicked Rotor (KR) model \cite{I90}, which consists of the free propagating 
rotor driven periodically in time by an external force. Since the total energy of a driven
system is not a constant of motion any more, the chaotic behavior can appear even in the 
one-dimensional ($1D$) case. 

Although many results are known for the standard $1D$ KR there is almost no study for its
two dimensional generalization, besides ref. \cite{DF88} where the authors have focused on
the analysis of dynamical evolution. The corresponding model is described by the following
Hamiltonian:
\begin{eqnarray}
\label{ham}
H &=& H_0 +  k V(\{\theta_i\})\sum_m \delta (t-mT), \\ 
H_0(\{{\cal L}_i\}) &=& \sum_{i=1}^2\frac{\tau_i}{2} ({\cal L}_i+\gamma_i)^2,\nonumber
\end{eqnarray}
where the index $i=1 (2)$ is related to the first (second) rotor correspondingly. ${\cal L}_i$ 
denotes the angular momentum and $\theta_i$ the conjugate angle of one rotor. The kick period 
is $T$, $k$ is the kicking strength, while $\tau$ is a constant inversely proportional to the 
moment of inertia of the rotor. The parameter $\gamma$ is an irrational number whose meaning 
will be explained below. The Hamiltonian (\ref{ham}) describes a system which is kicked 
periodically in time. Another representation of the Hamiltonian (\ref{ham}) may be given by 
one rotor moving on a two-dimensional sphere. The two rotors are interacting with each other 
by the potential
\begin{equation}
\label{pot-2d}
V(\{\theta_i\}) =\( \cos(\theta_1) \cos(\theta_2) \cos({\alpha}) +
{1\over2} \sin(2\theta_1) \cos(2\theta_2) \sin({\alpha})\)
\end{equation} 
The parameter $\alpha$ breaks time reversal symmetry (TRS). The form (\ref{pot-2d}) was chosen
so as to minimize the effect of symmetry breaking on the classical diffusion constant $D$ (see 
bellow).

\begin{figure}[!t]
\includegraphics[scale=1]{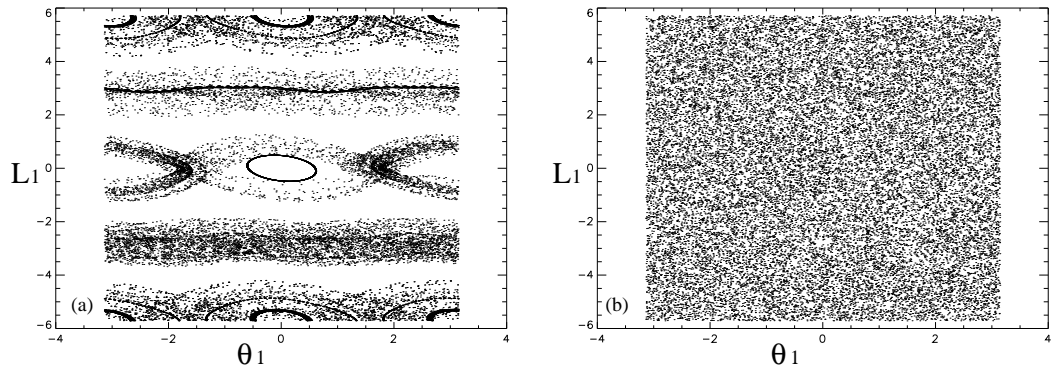}
\caption{\label{fig1}
Poincar\'e section of the classical phase space for Hamiltonian 
(\ref{ham}) for (a) $k=0.36$, (b) $k=6.37$}
\end{figure}   

\begin{figure}[!t]
\includegraphics[scale=0.5]{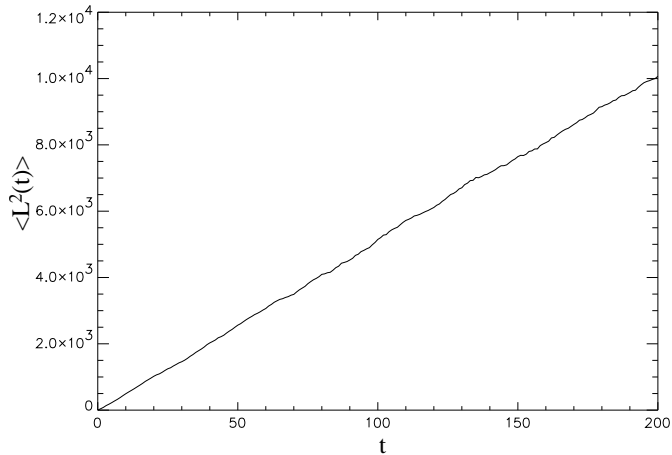}
\caption{\label{fig2}
Diffusion in momentum space for $k=9.1$}
\end{figure}

The classical equations of motion which correspond to the Hamiltonian (\ref{ham}) can be 
integrated over the kick period $T$ giving a set of simple maps:  
\begin{eqnarray}
\label{kr-maps2}
\thf &=& \theta_1(n)+\tau_1 T {\cal L}_1(n) \;\; \mbox{mod} \;\; 2 \pi \nonumber \\
\ths &=& \theta_2(n)+\tau_2 T {\cal L}_2(n) \;\; \mbox{mod} \;\; 2 \pi \nonumber \\
{\cal L}_1(n+1) &=& {\cal L}_1(n)-k(-\sin(\thf)\cos(\ths)\cos(\alpha)+\nonumber\\
&& +\cos(2\thf)\cos(2\ths)\sin(\alpha))  \nonumber \\
{\cal L}_2(n+1) &=& {\cal L}_2(n)-k(-\cos(\thf)\sin(\ths)\cos(\alpha)-\nonumber\\
&& -\sin(2\thf)\sin(2\ths)\sin(\alpha)) 
\end{eqnarray}
where $\theta_i(n)$ and ${\cal L}_i(n)$ are the values of the dynamical variables taken 
just after the $n$-th kick. The motion generated by this set of maps  is integrable in 
the absence of the kicking potential. For sufficiently small but non-zero $k$ the phase 
space of this system contains both regular islands and chaotic sea (see Fig~\ref{fig1}a). 
When $k$ is large enough then the dynamics becomes fully chaotic (see Fig.~\ref{fig1}b) 
and there is diffusion in momentum space (Fig.~\ref{fig2}) with diffusion coefficient 
\begin{equation}
D\equiv lim_{t \rightarrow \infty }{<{\bf{\cal L}}^2(t)>\over t} \simeq {k^2\over 2T}
\end{equation}
The last expression is correct within the random phase approximation \cite{DF88,I90,ZEN97} 
(see Appendix~II).

\begin{figure}[!t]
\includegraphics[scale=0.5]{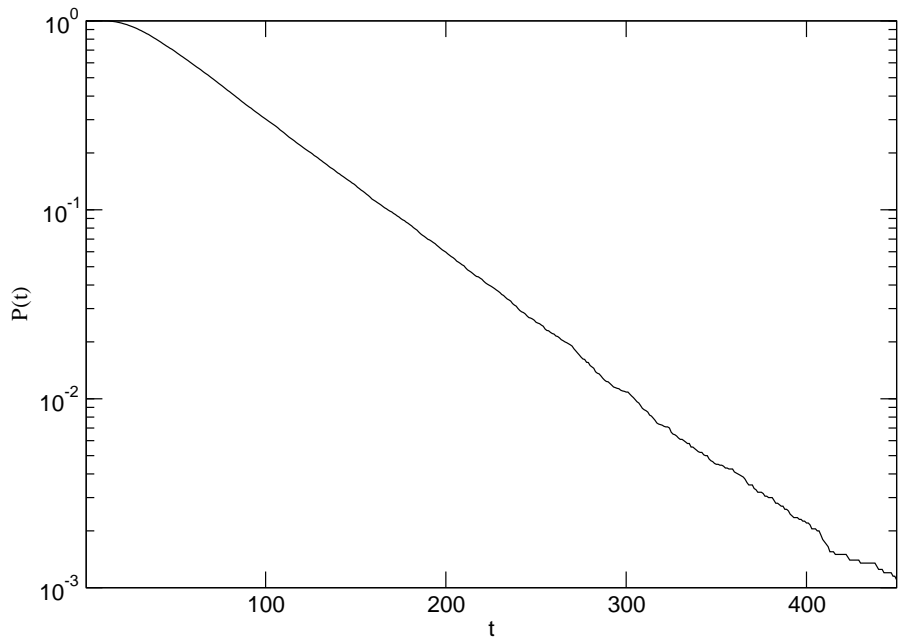}
\caption {\label{fig3} Exponential decay of the classical survival probability for open 
kicked rotor model.}
\end{figure}

The corresponding open system is described by the same set of equations (\ref{kr-maps2}) 
as the closed one, on top of which we add  absorbing boundary conditions. Namely, we set 
${\cal L}_i$ equal to zero, if ${\cal L}_i<0$ or ${\cal L}_i>L$. These conditions give to 
"the particles" the possibility to escape from the system. The evolution of the classical 
density $\rho (x,y,t)$ is described by the Frobenius-Perron equation, which takes in our
case the form of the following diffusion equation:
\begin{equation}
\label{dif-kr}
\frac{\partial \rho}{\partial t} = \frac{D}{4}\Delta \rho
\end{equation}
with absorbing boundary condition
\begin{equation}
\label{bcond}
\vec{J}_n\equiv -\frac{D}{4}(\vec{\nabla}\rho)_n=-\Phi_n
\end{equation}
which sets the flux at the boundary $\vec{J}_n$ to be equal to the number of
particles $\Phi_n$ carried out under one iteration of the map. The solution 
of Eq.~(\ref{dif-kr}) and Eq.~(\ref{bcond}) can be represented as the superposition
 of the diffusive  eigenmodes $v_k(x,y)$:
\begin{equation}
\rho (x,y,t)=\sum_{k=1}^{\infty}c_k e^{-\gamma_k t}v_k(x,y) 
\end{equation}
where $\gamma_k$ are the  corresponding eigenvalues and $c_k$ are coefficients
determined by the initial condition. The asymptotic behavior of the density
is governed by the smallest eigenvalue $\gamma_1\equiv \Gamma_{\rm cl}$.
As a consequence one has an exponential
decay of the classical survival probability $P(t)=\int\int dx\:dy\: \rho(x,y,t)$:
\begin{equation}
P(t)\propto e^{-\Gamma_{cl} t},
\end{equation}
The classical decay rate $\Gamma_{cl}\sim D/L^2$  can be estimated as the inverse time 
needed for the particle to reach the boundary (Thouless time). The exact value of the 
classical decay rate can be obtained as the solution of the corresponding eigenvalue 
problem or from the numerical calculation of $P(t)$. In Fig.~\ref{fig3} we present
the results of our numerical calculations for $P(t)$ for some representative parameters.
      
\section{Quantization on a torus }

Because of the periodicity of the external force a time dependent solution of the KR 
model can be represented as a superposition of Floquet states \cite{Reichl}. These are 
the eigenstates of the evolution operator for one period (Floquet operator). For kicked
systems the interaction with external force is instantaneous and one can factorize the 
total Floquet operator into the product of the evolution operators corresponding to the 
free propagation and the interaction. Due to this fact the kicked systems are very convenient 
for numerical study. 

If the ${\cal L}_i$ are taken mod$(2\pi m_i/T\tau_i)$ where $m_i$ are integers,
Eq.~(\ref{ham}) defines a dynamical system on a torus. The quantum mechanics of
this system is described by a finite-dimensional time evolution operator for one
period
\begin{equation}
\label{Uop}
U=\exp\[-iH_0(\{{\cal L}_i\})T/2\] \exp\[-iV(\{\theta_i\})\]\exp\[-iH_0(\{{\cal L}_i\})T/2\] 
\end{equation}
where we put $\hbar=1$. In the above representation, $U$ is symmetric and describes the 
evolution in the middle of the rotations between two successive kicks. Upon quantization, 
additional symmetries associated with the
discreteness of the momentum show up, which can be destroyed by introducing irrational
values for the parameters $\gamma_i$'s. The most striking consequence of quantization
is the suppression of classical diffusion in momentum space due to quantum dynamical
localization \cite{FGP82, I90}. We introduce the eigenstate components ${\bf \Psi}_k
(n)$ of the Floquet operator in the momentum representation by
\begin{equation}
\label{eigen}
\sum_m U_{mn} {\bf \Psi}_k(n) = e^{i\omega_k T} {\bf \Psi}_k(n) \quad .
\end{equation}
The quantities $\omega_k$ are known as quasi-energies, and their density is $\rho = 
T/2\pi$. The corresponding mean quasi-energy spacing is $\Delta = 1/(\rho L^d)$, where 
$L$ is the linear size of the system and $d$ is the dimensionality ($d=2$ in our case). 
The Heisenberg time is $t_H=2\pi/\Delta$ while $t_D=L^2/D$ is the diffusion time (Thouless 
time). Now one can formally define a dimensionless conductance as 
\begin{equation}
\label{tcond}
g=t_H/t_D = D_{k} L^{d-2}
\end{equation}
where $D_k=T D$ is the diffusion coefficient measured in number of kicks. Four length 
scales are important here: the wavelength $\lambda$, the mean free path $l_M$, the 
linear extent of the system $L$, and the localization length $\xi$. In the rest of this
paper we will always assume that
\begin{equation}
\label{ftpr}
\lambda \ll l_M \ll L \ll \xi.
\end{equation}
The first condition ensures that transport between scattering events may be treated
semiclassically. This limit can be achieved for our system (\ref{ham}) when $k
\rightarrow \infty$, $T\rightarrow 0$ while the classical parameter $K=kT$ remains
constant. When $l_M \ll L $ as long as the motion is not localized (i.e. $ L \ll
\xi$) it is diffusive, since a particle scatters many times before it can traverse
the system. The resulting mean free path for our system (\ref{ham}) is $l_M\simeq
{\sqrt D_k}$ (see Appendix I) while the localization length for 
$d=1$ is $\xi \simeq D_k/2$ \cite{I90} and for $d=2$ is $\xi \simeq l_M e^{D_k/2}$ 
\cite{DF88,Lee85}.

A great advantage of the kicked rotor consists in the fact that due to the unitarity of 
the evolution operator all its eigenstates have the same statistical properties. This is 
in contrast to the eigenstates of Hamiltonian models (like Anderson model), where the 
eigenstates belonging to different parts of the spectrum have different statistical 
properties and therefore they must be picked up from a narrow energy window. This allow
us to collect a huge set of numerical data and perform a rather accurate numerical
analysis. The results of our investigation will be discussed in section V. 

\section{The scattering Approach}

To proceed with the analysis of the resonance widths and delay times we turn the closed 
$2D$ KR model (\ref{Uop}) into an open one. To this end we impose `absorption' at the 
boundary of a square sample of size $L\times L$ in the momentum space in complete analogy
with the classical set-up. In other words, every time that one of the components of the 
two dimensional momentum $({\cal L}_1, {\cal L}_2)$ takes on the value $1$ or $L$, the 
particle leaves the sample. The corresponding unitary scattering matrix $S$ (see Appendix III) 
is given \cite{FS00}
\begin{equation}
\label{appSmatr}
S(\omega)=\sqrt{I-WW^{\dagger}}-WU\frac{1}{e^{-i\omega}-\sqrt{I-W^{\dagger} W} U}W^{\dagger}
\end{equation}
where $I$ is the $L^2\times L^2$ unit matrix and $W$ is a $M \times L^2$ matrix with 
matrix elements
\begin{equation}
\label{wmatrix}
W_{i,j}(k,A) =\left\{
\begin{array}{cc}
w_i, & i=j \\ \\
0, & i\neq j \
\end{array}
\right.   \nonumber
\end{equation}
with $w_i^2 \in \[0,1\]$ the tunnel probability in mode $i$. In the case of perfect coupling,
which is considered here, $w_i=1$. Then $W W^{\dagger} = I_{M\times M}$ and $W^{\dagger} W$ 
is a $L^2 \times L^2$ diagonal matrix with $M$ non-zero elements equal to one. From the 
physical point of view, $W$ describe at which "site" of the $L\times L$ sample we attach 
$M$ "leads" [in our case $M=4(L-1)]$. 
Here $W^{ \dagger}W$ is a projection operator onto the boundary, while $P\equiv I-
W^{\dagger} W$ is the complementary projection operator which satisfies
\begin{equation} 
\sqrt{I-W^{\dagger} W} = I-W^{\dagger} W
\end{equation}
Taking this into account, the expression (\ref{appSmatr}) can be simplified:
\begin{equation}
\label{smatrns}
S(\omega)=-WU\frac{1}{e^{-i\omega}-(I-W^{\dagger} W)U}W^{\dagger} = 
WU\frac{1}{U-e^{-i\omega}-W^{\dagger} WU}W^{\dagger}.  
\end{equation}
and using the unitarity of the evolution operator $U$ we can rewrite the last expression 
as follows:
\begin{equation}
\label{Smatrwig}
S=W\frac{1}{I-W^{\dagger} W-U^{\dagger}e^{-i\omega}}W^{\dagger}.
\end{equation}
The scattering matrix (\ref{Smatrwig}) can equivalently be written in the form used 
conventionally in quantum chaotic scattering:
\begin{equation}
\label{smatrix}
S(\omega) =  -W U e^{i\omega} {1\over I-e^{i\omega} P U} W^{\dagger} 
\end{equation}
The scattering matrix $S_{ij}$ given by Eq.~(\ref{smatrix}) can be interpreted in 
the following way: once a wave enters the sample, it undergoes multiple scattering 
induced by $[I-e^{i\omega} P U]^{-1}=\sum_{n=0}^{\infty} \left( e^{i\omega} PU\right)^n$ 
until it is transmitted out. It is clear therefore that the matrix ${\tilde U}=P U$ 
propagates the wave inside the sample. However, contrary to the closed system in which 
the evolution operator is unitary, the absorption breaks the unitarity of the evolution 
matrix ${\tilde U}$. 

The eigenvalues of ${\tilde U}$ occurring at complex quasi-energies $\tilde{\omega}_n 
= \omega_n - \frac{i}{2}\Gamma_n$ are the poles of the scattering matrix. They represent 
long-lived intermediate states to which bound states of a closed system are converted 
due to coupling to continua. Here $\omega_n$ and $\Gamma_n$ are the (dimensionless) 
position and width of the resonances, respectively. 

Having at our disposal the scattering matrix $S$ we can calculate the Wigner-Smith 
delay time. It captures the time-dependent aspects of quantum scattering and formally 
it is defined as
\begin{equation}
\label{wdt}
\tau_W (\omega) = -i {d\over d\omega} \ln \det S(\omega) 
\end{equation}
and can be interpreted as the typical time an almost monochromatic wave packet remains 
in the interaction region. 

A generalization of the  notion of Wigner delay time is given through the Wigner-Smith 
operator. The latter is defined as (for the kicked rotor model one should use quasi-energy 
instead of energy in the definition of Wigner-Smith operator) 
\begin{equation}
Q=\frac{1}{i}S^{\dagger}\frac{\partial S}{\partial \omega}
\end{equation}
Introducing a new operator 
\begin{equation}
K\equiv\frac{1}{I-W^{\dagger} W-U^{\dagger}e^{-i\omega}}  
\end{equation}
and taking the derivative of  both sides of Eq.~(\ref{Smatrwig}) we obtain
\begin{eqnarray}
\frac{\partial S}{\partial \omega}&=&W\frac{\partial K}{\partial \omega}
W^{\dagger}
=W\frac{1}{I-W^{\dagger} W-U^{\dagger}e^{-i\omega}}(-i)\wexp
U^{\dagger}\frac{1}{I-W^{\dagger} W-U^{\dagger}e^{-i\omega}}W^{\dagger}=\nonumber \\
&=&-i\wexp W K U^{\dagger} K 
W^{\dagger}
\end{eqnarray}
Then the definition of the Wigner-Smith yields
\begin{equation}
\label{qmatrix}
Q=-i\wexp W K^{\dagger} W^{\dagger} W K U^{\dagger} K W^{\dagger}
\end{equation}
The Wigner delay time can be expressed as the sum of proper delay times $\tau_q$. The
latter are the eigenvalues of the Wigner-Smith operator (\ref{qmatrix}).

\section{Distribution of eigenfunction intensities}

The statistical properties of wavefunction intensities have sparked a great deal of
research activity in recent years. These studies are not only relevant for mesoscopic
physics \cite{AKL91,MK95,FM94,FM95,FE95,FE95-2,M97,Mir00,M00-2,SA97,MKW90,UMRS00,UMS01,N01,
ARS02}, but also for understanding phenomena in areas of physics, ranging from nuclear 
\cite{ZBFH96} and atomic \cite{BUF96,BCS97} to microwave physics \cite{SS90,S91,KKS95,PS00} 
and optics \cite{NS97}. Experimentally, using microwave cavity technics it is possible to 
probe the microscopic structure of electromagnetic wave amplitudes in chaotic or disordered
cavities \cite{SS90,S91,KKS95,PS00}. Recently, the interest in this problem was renewed when
new effective field theoretical techniques were developed for the study of the
distribution of eigenfunction intensities ${\cal P} (|\psi|^2)$ of {\it random}
Hamiltonians. As the disorder increases, these results predict that, the eigenfunctions 
become increasingly non-uniform, leading to an enhanced probability of finding
anomalously large eigenfunction intensities in comparison with the random matrix 
theory prediction. Thus, the notion of {\it pre-localized} states, i.e. states which are
localized much more strongly than  typical eigenstates, has been introduced
\cite{AKL91,MK95,FM94,FM95,FE95,FE95-2} to explain the appearance of long tails in the
distributions of the conductance and other physical observables \cite{AKL91}.

\begin{figure}[!t]
\includegraphics[scale=0.5]{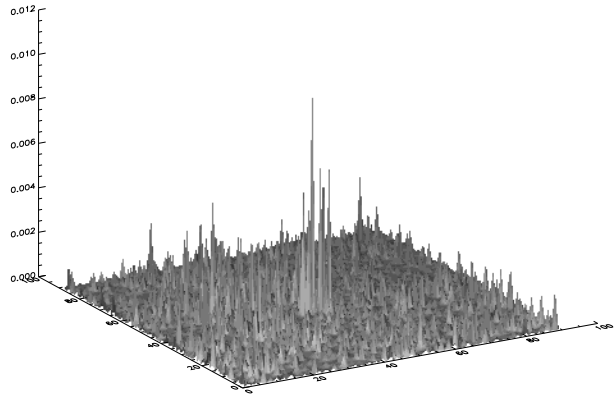}
\caption {\label{single-als}An example of anomalously localized state. The size of the 
system is $L\times L = 90\times 90$, the diffusion coefficient $D=33.8$}
\end{figure}

Up to now all theoretical predictions \cite{AKL91,MK95,FM94,FM95,FE95,FE95-2,M97,Mir00,
M00-2,SA97,ARS02} and numerical calculations \cite{UMRS00,UMS01,N01} apply to disordered 
systems and are based on an ensemble averaging over disorder realizations. Their validity, 
however, for a quantum {\it dynamical} system (with a well defined classical limit) that
behaves diffusively is not evident. Here we show that pre-localized states exist
also for dynamical systems with underlying classical diffusion and investigate the 
effect of these states in ${\cal P}(|\psi|^2)$. An example of them is reported in 
Fig.~(\ref{single-als}). Our main conclusion is that in a generic dynamical system 
with classical diffusion, ${\cal P}(|\psi|^2)$ is described quite well by the 
nonlinear $\sigma-$model (NLSM). We point out  that between the various theoretical
works there is a considerable disagreement about the parameters that control the shape 
of ${\cal P}(|\psi|^2)$ and their dependence on time-reversal symmetry (TRS). More 
specifically, the NLSM suggests that the tail of  ${\cal P}(|\psi|^2)$ in two dimensions 
($2d$) is sensitive to TRS \cite{FE95,FE95-2,M97,Mir00,M00-2}, while a direct optimal
fluctuation (DOF) method predicts a symmetry independent result \cite{SA97}. This 
prediction was an additional motivation for the present study.

In the next subsection we calculate the distribution function ${\cal P}(y=L^d |{\bf 
\Psi}_k(n)|^2)$ by using a direct diagonalization of the Floquet operator (\ref{Uop}). 
The TRS is broken entirely for $\alpha=5.749$, i.e.  the phase difference for a typical 
trajectory and its time-reversal counterpart is larger than $2\pi$, so that all interference
effects between time-reversal trajectories are suppressed. In order to test the issue of dynamical 
correlations, we randomize the phases of the kinetic term of the evolution operator 
(\ref{Uop}) and calculate the resulting ${\cal P}(y)$. This model will be referred 
to as Random Phase KR (RPKR). Since all our eigenfunctions have the same statistical 
properties (in contrast to the Anderson cases where one should pick up only eigenfunctions 
having eigenenergies 
within a small energy interval \cite{UMRS00,UMS01,N01}) we make use of all of them in 
our statistical analysis. The classical parameter  $K>6.36$ is large enough in all cases to 
exclude the existence of any stability islands in phase space, the accelerator modes
known for KR \cite{ZEN97, IZ02} are  avoided as well. The influence of the accelerator modes 
on the eigenfunction statistics is an interesting problem, which is out of the scope of the
 present study and deserves a separate investigation. The classical diffusion 
coefficient $D_k$ is calculated numerically by iterating the classical map obtained from 
(\ref{ham}). Below we present our numerical results and compare them to the predictions 
of Refs.~\cite{MK95,FM94,FM95,FE95,FE95-2,M97,Mir00,M00-2,SA97}.

\subsection{Numerical results}

\begin{figure}[!t]
\includegraphics[scale=0.5]{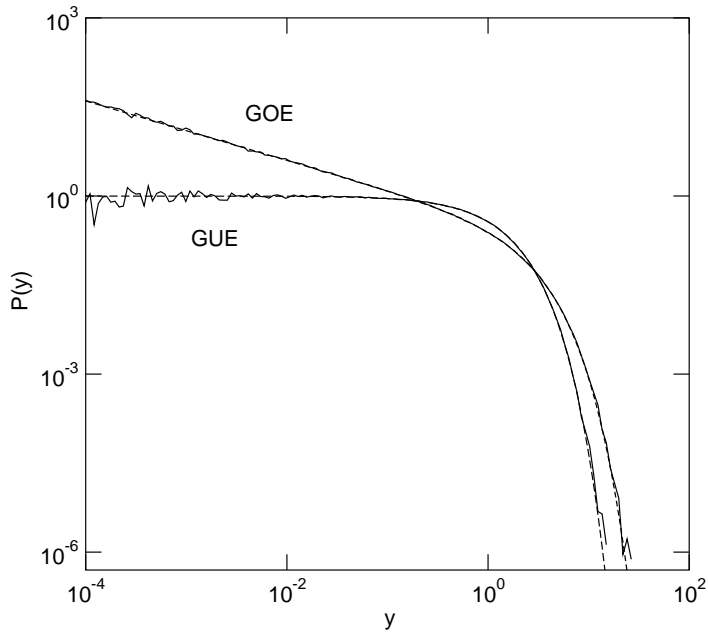}
\caption {\label{wf-rmt} Distributions of the eigenfunction intensities in the ballistic
regime for two-dimensional kicked rotor (solid lines) compared with the 
RMT predictions (dashed lines).}
\end{figure}

Let us start our numerical analysis from the ballistic regime where $g \rightarrow 
\infty$. In this case, RMT is applicable and one finds \cite{S99}:
\begin{eqnarray}
\label{wf-rmt-goe}
{\cal P}_{(\beta=1)}^{RMT}(y)&=& \exp(-y/2)/\sqrt{2\pi y}\\
\label{wf-rmt-gue} 
{\cal P}_{(\beta=2)}^{RMT}(y) &=& \exp(-y).
\end{eqnarray}
Here $\beta$ denotes the corresponding Dyson ensemble: $\beta=1(2)$ for preserved 
(broken) TRS. This result can be easily understood. Indeed, within the  random matrix 
theory one assumes that all the eigenvector components are independent (the normalization 
of the eigenvector is not essential in the thermodynamic limit, i.e. when the number 
of its components becomes very large) random variables obeying Gaussian distribution. 
Going to the distribution of the modulus square of the components one immediately 
recovers Eq.~(\ref{wf-rmt-goe}). For the case of the broken time reversal symmetry
one should take into account that each component has statistically independent real 
and imaginary parts, leading to the distribution given by Eq.~(\ref{wf-rmt-gue}). The 
numerical data presented in  Fig.~(\ref{wf-rmt}) shows that the distributions of the 
eigenfunction intensities in the ballistic regime for the two-dimensional kicked rotor 
are described very nice by the RMT prediction.
 
As the ratio between the system size and the localization length increases, 
the deviations from the RMT results of the body and the tails 
of the distribution ${\cal P }_{\beta}(y)$ become noticeable and can be parameterized by 
a single parameter which is the dimensionless conductance $g=D_k$.
For $y< D_k$, according to all studies \cite{FM94,FM95,FE95,FE95-2,M97} ${\cal P}
(y)$ is just the RMT result with polynomial corrections in powers of $L/D_k$, i.e. $
{\cal P}_{\beta}(y)= {\cal P}_{\beta}^{RMT}(y)[1+ \delta {\cal P}_\beta(y)]$. The leading 
term of this expansion is given by
\begin{eqnarray}
\label{q1dor}
\delta {\cal P}_{\beta}(y)  &\simeq & \kappa
\left\{
\begin{array}{ll}
3/4-3y/2+y^2/4 &\mbox{for$\,\beta=1$}\,\\
1-2 y+y^2/2    &\mbox{for$\,\beta=2$}\,
\end{array}
\right. ,
\end{eqnarray}
where $\kappa$ is the $2d$ diffusion propagator (time-integrated return probability),
 which is identical for $\beta=1$ and $\beta=2$ since it is a classical quantity.

\begin{figure}[!t]
\includegraphics[scale=0.5]{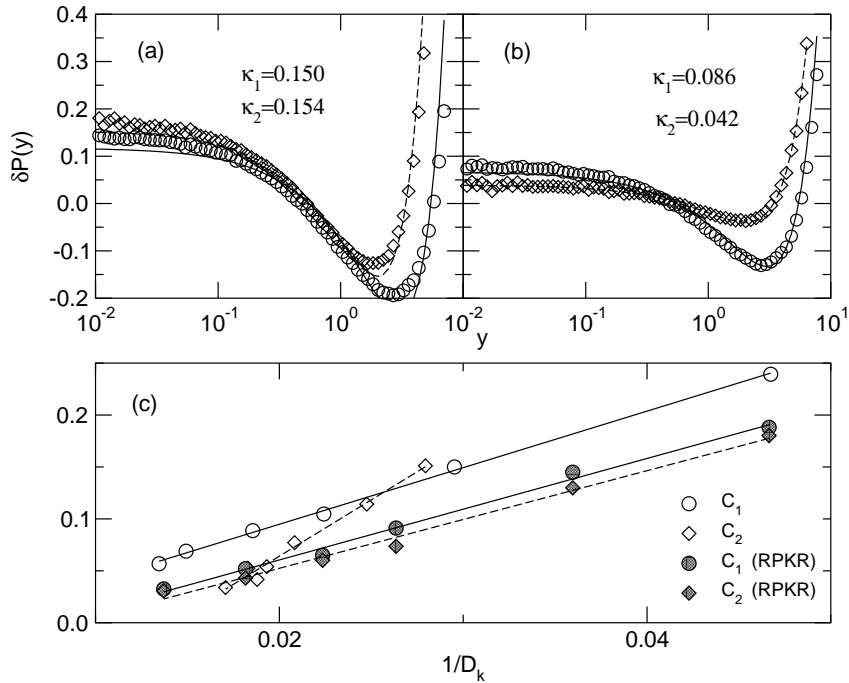}
\caption {\label{fig6} 
Corrections to the distribution intensities $\delta {\cal P}_{\beta} (y)$ for the two-
dimensional kicked rotator model (\ref{Uop}). The system size is $L=90$, ($\circ$) $
\beta=1$, ($\diamond$) $\beta=2$. The solid (dashed) lines are the best fit of (\ref{q1dor}) 
for $\beta=1 (2)$ to the numerical data: (a) $D_k\approx 34$ and (b) $D_k\approx 53$ ; 
(c) Fit parameters $\kappa_{\beta}$ vs. $D_k^{-1}$. The solid (dashed) lines are the best 
fits to $\kappa_{\beta}= A_{\beta} D_k^{-1}+ B_{\beta}$ for $\beta=1 (2)$.}
\end{figure}

Figures \ref{fig6}(a) and (b) show corrections to ${\cal P}_{\beta}^{RMT}$ for $g=
D_k\gg 1$ for two representative values of $D_k$. We find again that the form of the 
deviations are very well described by Eq.~(\ref{q1dor}) and the agreement becomes better 
for larger values of the diffusion constant. This is due to the fact that by increasing 
$D_k$ we are approaching the semiclassical region and therefore Eqs.~(\ref{ftpr}) are 
better satisfied. At the same time higher order corrections in $\delta {\cal P}_{\beta}
(y)$ become negligible with respect to the leading term given by Eq.~(\ref{q1dor}). 

In Fig.~\ref{fig6}(c) we summarize our 
results for various $D_k$ values. The extracted $\kappa_{\beta}$ values are obtained 
by the best fit of the data to Eq.~(\ref{q1dor}). Again we find that $\kappa_{\beta}$ 
depends linearly on $1/D_k$.  However, $\kappa_1$ and $\kappa_2$, are different. Moreover 
the best fit with $\kappa_{\beta}=A_{\beta}D_k^{-1}+ B_{\beta}$ yields $A_{\beta=1}=5.44
\pm0.03$ and $A_{\beta=2}=10.84\pm 0.04$ indicating
that the ratio $R=A_2/A_1$ is close to $2$, a value that could be explained on the basis
of ballistic effects \cite{Mir00,M00-2,UMRS00,UMS01}. Taking the latter into account 
leads to an additional term in the classical propagator $\kappa_{\beta}=\kappa_{diff}
+{\beta \over 2} \kappa_{ball}$. The first term is the one discussed previously and is
associated with long trajectories which are of diffusive nature while the latter one 
is associated with short ballistic trajectories which are self-tracing \cite{Mir00,M00-2}.
Thus, when $\kappa_{diff} \ll \kappa_{ball}$ we get $R=2$. The calculation with the 
RPKR model shows, however, that the corresponding ratio is $R\simeq 1$ in agreement
with the theoretical prediction for disordered systems with a pure diffusion. This
indicates that dynamical correlations can be important. 

For the tails of the distributions, the result of the NLSM within a saddle-point
approximation \cite{FE95,FE95-2,Mir00,M00-2} is
\begin{equation}
\label{nlsmt}
{\cal P}_\beta(y) \simeq \exp[-C_{\beta}^{\sigma } (\ln y)^2 ] , \quad
C_{\beta}^{\sigma} = {\beta\pi^2\rho\over2} {D\over\ln (L/ l)}.
\end{equation}
Note that the decay in the  tails of Eq.~(\ref{nlsmt}) depends on $\beta$. Recently, 
a DOF method was used to calculate the tails of ${\cal P}_{\beta}(y)$ \cite{SA97} for the
white-noise random potential. 
It was found that the tails are still given by Eq.~(\ref {nlsmt}) but with a log-normal 
coefficient $C$ which is independent of the parameter $\beta$ :
\begin{equation}
\label{ofm}
C^{\rm DOF} = \pi^2 \rho {D\over\mbox{ln}(L/\lambda)}.
\end{equation}

\begin{figure}[!t]
\includegraphics[scale=0.5]{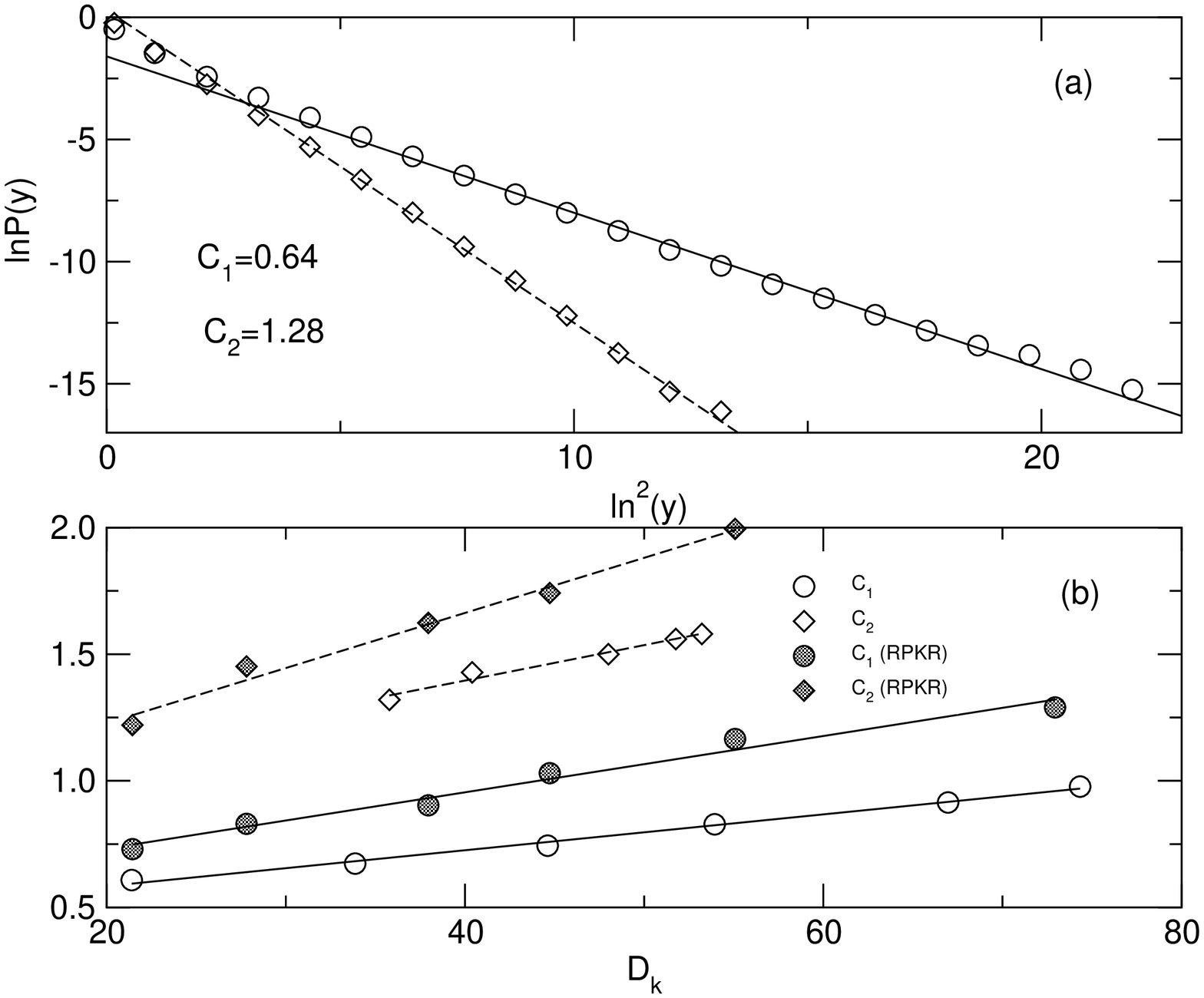}
\caption {\label{fig7} 
(a) Tails of the distribution ${\cal P}_{\beta=1}(y>D_k)$ for the model (\ref{Uop})
and $D_k\simeq 35$. The system size is $L=80$, ($\circ$) $\beta=1$, ($\diamond$) 
$\beta=2$. The solid (dashed) lines are the best fit of (\ref{nlsmt}) for $\beta=1 (2)$ 
to the numerical data; (b) Fitted log-normal coefficients $C_{\beta}$ versus the
classical diffusion coefficient $D_k$. The solid (dashed) lines are the best fits
to $C_{\beta}= A_{\beta} D_k+ B_{\beta}$ for $\beta=1 (2)$. }
\end{figure}

Figure \ref{fig7}(a) shows a representative case of ${\cal P}_{\beta=1}(y> D_k)$. The 
tails show a log-normal behavior predicted by Eq.~(\ref{nlsmt}). In Fig.~\ref{fig7}(b) 
we report the log-normal coefficients $C_{\beta}$ extracted from the best fit to our
numerical data, versus the classical diffusion coefficient. A pronounced linear behavior
is observed in agreement with both theories. However one clearly sees that $C_1$
differs from $C_2$ in contrast to the DOF prediction (\ref{ofm}) and to recent numerical
calculations done for the $2d$ Anderson model \cite{UMRS00}. We point out here that in
\cite{UMRS00} the authors were not able to go to large enough values of conductance
$g$ (in comparison to our study) where the theory can really be tested. In contrast,
the NLSM predicts a value of $2$ for the ratio $R=C_2^{\sigma}/C_1^{\sigma}$. We note
that $C_{\beta}^{\sigma}$ is only the leading term in $D_k$. In order to calculate
this ratio, we performed a fit to our data with $C_{\beta}=A_{\beta} D_k+ B_{\beta}$.
The resulting ratio was found to be $R=A_2/ A_1= 1.97 \pm 0.03$ in perfect agreement
with the NLSM predictions. Finally in Fig.~\ref{fig7}(b) we also present our results 
for the RPKR model (using the same data as the one in Fig.~\ref{fig6}(c)). Again we 
found that the ratio $R= 1.96\pm 0.03 \approx 2$. Thus ${\cal P}(y> D_k)$ depends on TRS 
and is described by the NLSM. The fact that the prediction of a DOF method is not observed 
in our calculations might be due to non-universal (depending on the type of disorder) character 
of this result.

\section{Open systems: distribution of resonances and delay times}
In this section we analyze the statistical properties of resonances ${\cal P}(\Gamma)$
and Wigner delay times ${\cal P}(\tau)$ for a $2D$ diffusive system. This study is 
important for various applications among which are diffusive random lasers and microwave 
cavities \cite{KM00,BB01,SS03,DSF90} where most of the theoretical treatment is limited by RMT. We 
point here that current developments of microwave experiments in random dielectric 
media in the diffusive regime \cite{CZG03} may allow us a direct comparison between 
theory and experiment.

Specifically, we have found that the resonance width distribution ${\cal P} (\Gamma)$ is given by
\begin{eqnarray}
\label{gammas}
{\cal P}(\Gamma < \Gamma_{cl}) & \sim & \exp(-C_\beta (\ln\Gamma)^2)\,, 
\,\,\,\,\,{\rm where}\,\,\,\,\,
C_{\beta}\sim \beta D \nonumber\\
{\cal P}(\Gamma\gtrsim \Gamma_{cl}) & \sim & {\sqrt {D\over L^2}}{1\over \Gamma^{3/2}}
\,\,\,\,\,\,\,\,
\end{eqnarray}
while the distribution of the Wigner delay times, is given by the following expressions:
\begin{eqnarray}
\label{taus}
{\cal P}(\tau\lesssim \Gamma_{cl}^{-1}) & \sim & {1\over\tau^{3/2}}\exp(-\sigma/\tau)\,\,\,\,\,\,\,
\,\,\,\, \,\, \nonumber\\
{\cal P}(\tau > \Gamma_{cl}^{-1}) & \sim & \exp(-C_\beta (\ln\tau)^2)\,\,\,
\end{eqnarray}
where $D$ is the classical diffusion constant,  $\beta$ denotes the symmetry class and 
$\sigma$ is some constant of order unity.  

Our theoretical
considerations are supported by numerical calculations for the open analog of the  $2D$ KR model 
described in sections II and  IV. The parameters of the model were chosen in such a way that the conditions 
(\ref{ftpr}) discussed in the previous section were fulfilled. In order to improve our 
statistics, we randomized the phases of the kinetic term of the evolution operator 
(\ref{Uop}) and used a number of different realizations. The results of the previous section
 allow us to conclude that this procedure doesn't change the universal features of the model. In 
all cases we had at least 60000 data for statistical processing. 

\subsection{Resonance widths distribution}
\label{sec-rwd}

For diffusive mesoscopic samples, there is no systematic investigation of ${\cal P}
(\Gamma)$ besides Ref.~\cite{BGS91} where the authors have focused on the large tails 
of ${\cal P}(\Gamma)$ for a quasi-1D system in the diffusive regime. This deficiency 
is felt especially strong in the random-laser community where one wants to know the 
statistical properties of the lasing threshold.

Usually we model random laser as a disordered or chaotic system containing a dye that 
is able to amplify the radiation with a rate $\eta$, in a certain frequency interval. 
In contrast to the traditional lasers where the necessary feedback is due to mirrors at 
the boundaries of the laser cavity, the key mechanism for random lasers is the multiple 
scattering inside the medium \cite{WAL95}. The lasing threshold is given by the value 
of the smallest decay rate (i.e. smallest resonance width) of all eigenmodes in the 
amplification window \cite{MB98,PB99}. The underlying reasoning is that in the mode 
with the smallest decay rate the photons are created faster by amplification than they 
can leave (decay) the sample.

Assuming that the number of modes $K\gg 1$ that lies in the frequency window where the 
amplification is possible, have resonance widths $\Gamma$ that are statistically 
independent one gets for distribution of lasing thresholds ${\tilde {\cal P}} (\Gamma)$ 
\cite{P03,PB99,MB98}:
\begin{equation}
\label{laserr}
{\tilde {\cal P}} (\Gamma) = K {\cal P}(\Gamma<<1) \left[ 1- \int_0^{\Gamma} {\cal P}
(\Gamma'<<1) d\Gamma'\right]^{K-1}
\end{equation}
where we have assumed that all $K$ resonances are distributed according to ${\cal P}
(\Gamma<<1)$. The validity of this approximation was verified recently in the framework of
the RMT \cite{FM02}. An important outcome of our study will be that one can identify traces
of pre-localized states in the latter distribution and consequently in ${\tilde {\cal P}} 
(\Gamma)$. This send some light to recent experimental finding for random lasers which 
suggests the appearance of localized modes in diffusive samples \cite{CLXB02}.

We start our analysis with the study of resonance 
width distribution ${\cal P}(\Gamma)$ for $\Gamma < \Gamma_{cl}$. 
The small resonances $\Gamma < \Delta$ can be associated, with the existence of 
pre-localized states of the closed system which were discussed in the previous
section.
 They consist of a short-scale bump (where most of 
the norm is concentrated) and they decay rapidly in a power law fashion from 
the center of localization \cite{Mir00,SA97}. One then expects that states of this 
type with localization centers at the bulk of the sample are affected very weakly by
the opening of the system at the boundaries. In first order perturbation theory,
considering the opening as a small perturbation we obtain
\begin{equation}
\label{pertgamma}
{\Gamma\over 2} = \langle \Psi|W^{\dagger}W|\Psi\rangle =\sum_{n\in {\rm boundary}}|\Psi(n)|^2 \sim 
L |\Psi(L)|^2 
\end{equation}
where $|\Psi (L)|^2$ is the wavefunction intensity of a pre-localized state at the 
boundary. At the same time the distribution of $\theta=1/{\sqrt L}\Psi (L)$ for large 
values of the argument is found to be of log-normal type \cite{SA97}:
\begin{equation}
{\cal P}(\theta) \sim \exp\left(-\pi^2D\ln^2 (\theta^2)\right)
\end{equation}
 Using this together with Eq.~(\ref{pertgamma}) we obtain 
\begin{equation}
{\cal P}(1/\Gamma) \sim \exp\left(-\pi^2D\ln^2 (1/\Gamma)\right)
\end{equation}
 We would like to stress that, based on the results of the previous section,
 the expression for ${\cal P}(\theta)$, must be corrected by including the TRS factor 
$\beta$ in the exponent.  Taking all the above into account we end up with the expression 
given in Eq.~(\ref{gammas}).  

\begin{figure}[!t]
\includegraphics[scale=0.5]{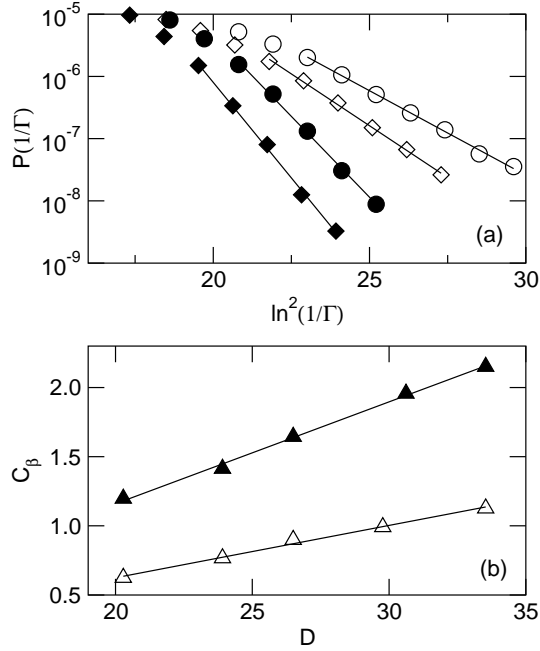}
\caption {\label{fig8} (a) The distribution of resonance widths (plotted as ${\cal P}(1/\Gamma)$
vs. $1/\Gamma$) for $\Gamma < \Gamma_{cl}$ for two representative values of $D$. The 
system size in all cases is $L=80$. Filled symbols correspond to broken TRS. 
The solid lines 
are the best fit of Eq.~(\ref{gammas}) for $\beta = 1 (2)$ to the numerical data. (b) 
Coefficients $C_{\beta}$ vs. $D$. The solid lines are the best fits to $C_{\beta}= 
A_{\beta} D+ B_{\beta}$ for $\beta=1 (2)$. The ratio $R=A_2/A_1 = 1.95\pm 0.03$} 
\end{figure}

The numerical data reported in Fig.~\ref{fig8} support the validity of
 the above considerations. 
However, we would like to mention that the perturbative argument is valid only for the 
case of very small resonances i.e. $\Gamma < \Delta$, whereas our numerical data 
indicate that one can extend the log-normal behavior of ${\cal P}(\Gamma)$ up to 
resonances with $\Delta < \Gamma < \Gamma_{cl} $. 

\begin{figure}[!t]
\includegraphics[scale=0.5]{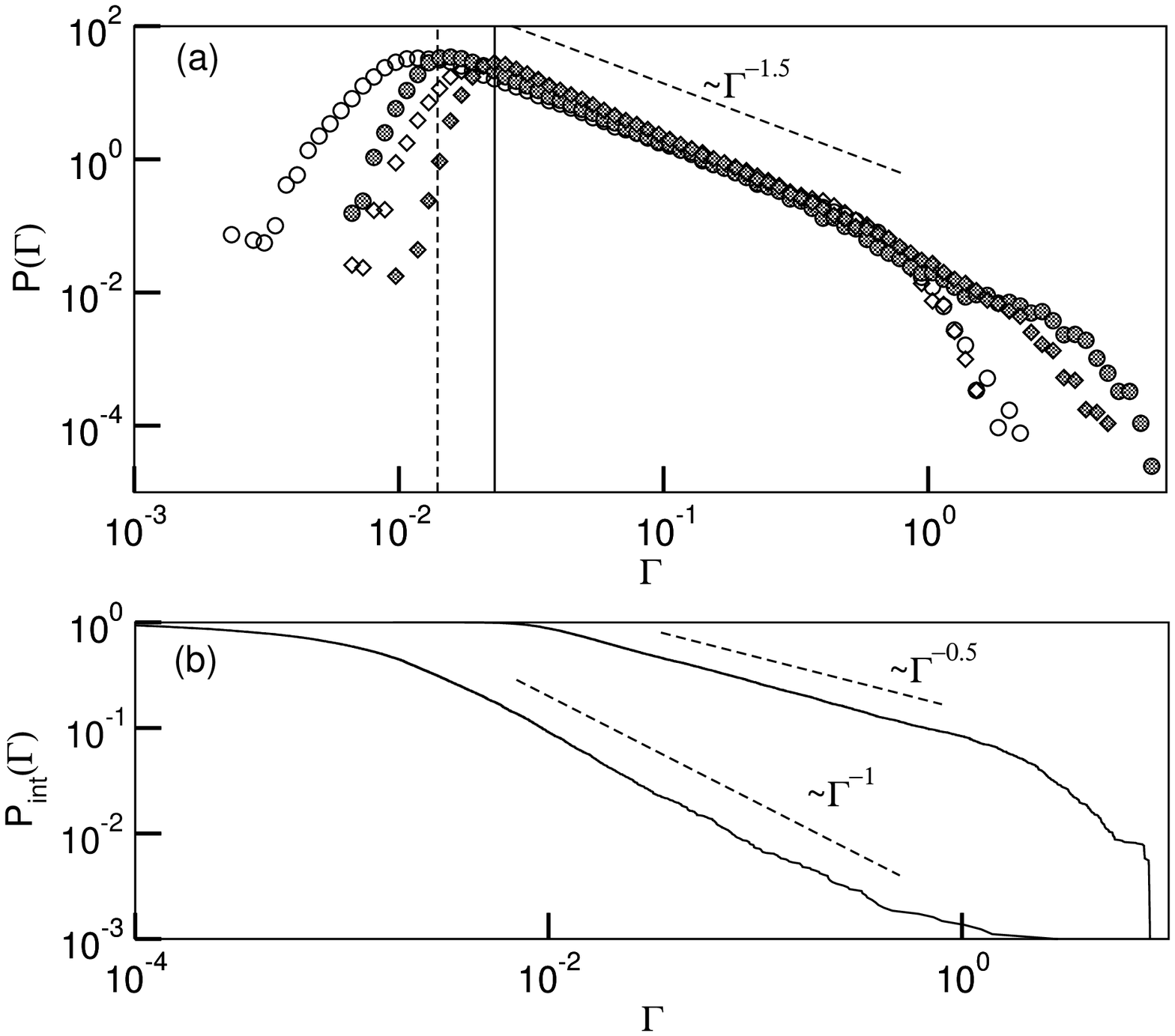}
\caption {\label{fig9}(a) The resonance width distribution ${\cal P}(\Gamma)$ for preserved
TRS and $D=20.3$ ($\circ$) and $D=33.5$ ($\diamond$). The corresponding filled symbols
represent ${\cal P}(\Gamma)$ for broken TRS and the same values of $D$. The dashed 
(solid) vertical line mark the classical decay rate $\Gamma_{cl}$  for 
$D=20.3 (D= 33.5)$. (b) The ${\cal P}_{\rm int} (\Gamma)$ for a sample with nine leads 
(lower curve). For comparison we plot also the ${\cal P}_{\rm int}(\Gamma)$ for the same 
sample but when we open the system from the boundaries. The dashed lines correspond 
to the theoretical predictions (\ref{obound}) and (\ref{final}).}
\end{figure}

Next we turn to the analysis of ${\cal P}(\Gamma)$ for $\Gamma \gtrsim \Gamma_{cl}$. In 
Fig.~\ref{fig9}(a) we report our numerical results for ${\cal P}(\Gamma)$ with preserved (broken) 
TRS for two representative values of $D$. An inverse power law ${\cal P}(\Gamma) 
\sim {\Gamma}^{-1.5}$ is evident in accordance with Eq.~(\ref{gammas}). 
(The behavior of the extreme large $\Gamma$ tails of ${\cal P}(\Gamma)$ 
is essentially determined by the coupling to the leads which is model dependent. Their 
relative number is proportional to $M/L^2 \sim L^{-1}$ and therefore they are 
statistically insignificant.) 
>From the figure it is clear that this part of the distribution is independent of
the symmetry class, in contrast to the small resonance distribution discussed above.

The following argument provides some understanding of the behavior of ${\cal P}(
\Gamma)$ for $\Gamma\gtrsim \Gamma_{cl}$. First we need to recall that the inverse of 
$\Gamma$ represents the quantum lifetime of a particle in the corresponding resonant 
state escaping into the leads. Moreover we assume that the particles are uniformly 
distributed inside the sample and diffuse until they reach the boundaries, where they 
are absorbed. Then we can associate the corresponding lifetimes with the time $t_R\sim 
1/ \Gamma_R \sim R^2/D$ a particle needs to reach the boundaries, when starting a 
distance $R$ away. This classical picture can be justified for all states with $\Gamma 
\gtrsim \Gamma_{cl} \sim D/L^2$. The relative number of states that require a time $t
<t_R$ in order to reach the boundaries (or equivalently the number of states with 
$\Gamma>\Gamma_R$) is 
\begin{equation}
\label{igam}
{\cal P}_{\rm int}(\Gamma_R)=\int_{\Gamma_R}^{\infty}{\cal P}(\Gamma)d\Gamma
\sim\frac{S(t_R)}{L^2}
\end{equation}
where $S(t_R)$ is the area populated by all particles with lifetimes $t<t_R$. In the 
case of open boundaries we get
\begin{equation}
\label{obound}
{\cal P}_{\rm int}(\Gamma_R)\sim \frac{L^2-(L-2R)^2} {L^2}\sim
\sqrt{\frac{\Gamma_{cl}}{\Gamma_R}}-\frac{\Gamma_{cl}}{\Gamma_R}.
\end{equation}
For $\Gamma_R > \Gamma_{cl}$ the first term in the above equation is the dominant one
and thus Eq.~(\ref{gammas}) follows.

Here it is interesting to point that a different way of opening the system might lead 
to a different power law behavior for ${\cal P}(\Gamma)$. Such a situation can be 
realized if instead of opening the system at the boundaries we introduce "one-site" 
absorber (or one "lead") somewhere in the sample. In such a case we have 
\begin{equation}
\label{final}
{\cal P}_{\rm int}(\Gamma_R)\sim\frac{S(t_R)}{L^2}=\frac{R^2}{L^2}=\frac{Dt_R}{L^2}\sim
\frac{\Gamma_{cl}}{\Gamma_R}.
\end{equation}
The above result is valid for any number $M$ of "leads" such that the ratio $M/L^2$ scales
as $1/L^2$. In Fig.~\ref{fig9}(b) we report the integrated resonance width distribution ${\cal 
P}_{\rm int}(\Gamma)$ for the case with nine "leads" attached somewhere to the $2D$ 
sample. 

A straightforward generalization of our arguments for 3D systems in the diffusive regime 
gives 
\begin{equation}
{\cal P}_{\rm int}(\Gamma_R) \sim \sqrt{\frac{\Gamma_{cl}}{\Gamma_R}} - 2 
\frac{\Gamma_{cl}}{\Gamma_R} +\frac{4}{3}
\left(\frac{\Gamma_{cl}}{\Gamma_R}\right)^{\frac{3}{2}}
\end{equation}
which for $\Gamma_R > \Gamma_{cl}$ leads to the same universal expression as in Eq.~(\ref{gammas}). 
Similarly, the analog of Eq.~(\ref{final}) in 3D is 
\begin{equation}
{\cal P}_{\rm int}(\Gamma_R)\sim 
\left(\frac{\Gamma_{cl}}{\Gamma_R}\right)^{\frac{3}{2}}.
\end{equation}

\begin{figure}[!t]
\includegraphics[scale=0.5]{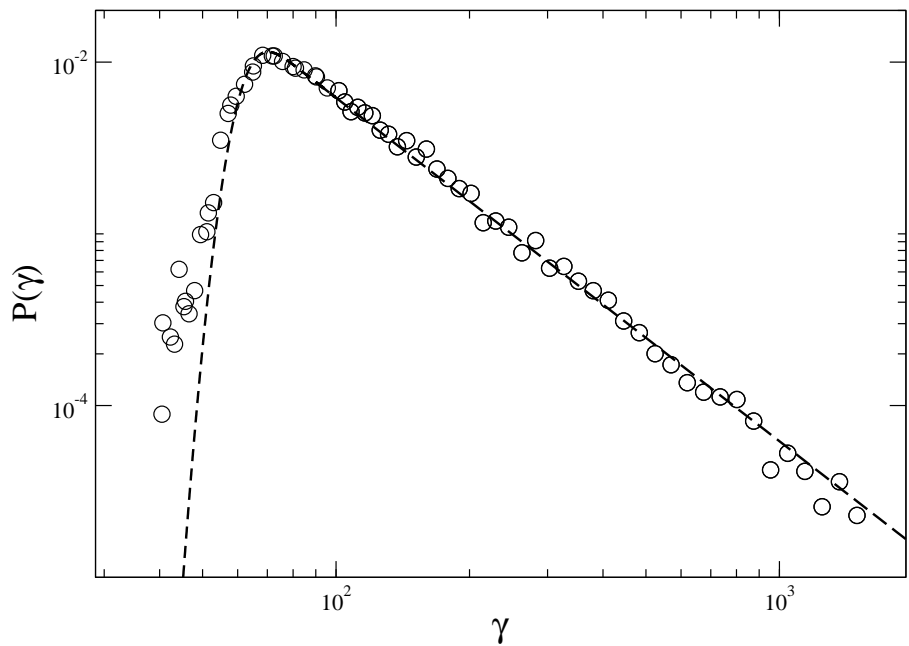}
\caption {\label{gbal} Distributions of the resonance widths in the ballistic regime 
for two-dimensional kicked rotor ($\circ$) compared with the RMT predictions (dashed 
lines).
}
\end{figure}

It is interesting to compare the above prediction (\ref{gammas}) with the results of 
the random matrix theory. In the general case, Fyodorov and Sommers \cite{FyodSom97} 
proved that the distribution of scaled resonance widths $\gamma=\Gamma/\Delta$ for 
the unitary random matrix ensemble, is given by
\begin{equation}
\label{FSpole}
{\cal P}(\gamma) = \frac {(-1)^M}{\Gamma(M)} \gamma^{M-1} {d^M\over 
d\gamma^M}
\left({\rm e}^  {-\gamma \pi q} {\sinh(\gamma\pi)\over(\gamma\pi)}\right)
\end{equation}
where $M$ is the number of open channels and the parameter $q$ controls the degree of
coupling with the channels ($q=1$ for perfect coupling). Our numerical data from the
$2D$ KR model in the ballistic regime, reported in Fig.~\ref{gbal} are in excellent
agreement with the theoretical prediction (\ref{FSpole}).

In the limit of $M\gg 1 $, which is relevant for the comparison with Eq.~(\ref{gammas}), 
Eq.~(\ref{FSpole}) reduces to the following expression \cite{FyodSom97}
\begin{equation}
\label{gam-rmt}
{\cal P}(\gamma)=
\left\{
\begin{array}{lll}
{M\over 2\pi \gamma^2}&\ {\rm, for}\ &{M\over \pi(q+1)}<
\gamma<{M\over\pi(q-1)} \\
\ 0&  {\rm, otherwise} &
\end{array}
\right.  \ .
\end{equation}
One can see two essential distinctions between this result and Eq.~(\ref{gammas}).
Firstly, the power law $P(\Gamma)\sim 1/\Gamma^2$ is not the same as the 
power law predicted by Eq.~(\ref{gammas}) for large resonances $P(\Gamma) 
\sim 1/\Gamma^{3/2}$. Since this difference appears in the "classical"
part of the distribution, it can be explained as a difference in the
classical dynamics of a particle inside the system: ballistic (RMT) versus
diffusive motion. Indeed, taking into account that for ballistic system $R\sim
v t$ and $\Gamma_{cl}\sim v/L$, where $v$ is the velocity of the particle,
one immediately finds from Eq.~(\ref{obound}) that ${\cal P}_{\rm int}(\Gamma_R)
\sim {\Gamma_{cl}}/{\Gamma_R}$ for $\Gamma_R >\Gamma_{cl}$, in agreement with
the RMT prediction $P(\Gamma)\sim 1/\Gamma^2$. Secondly, according to 
Eq.~(\ref{gam-rmt}) there is 
a gap in the distribution of the resonance widths: there are no resonances
with widths smaller than $M\over \pi(g+1)$. The existence of the gap can be understood,
if one relates the small resonances to the coupling of the wavefunctions to the leads.
Since the wavefunctions in the RMT are extended, the probability to find a wavefunction,
which is weakly coupled to all $M$ channels, goes to zero when the number of channels
becomes very large $M \gg 1$. In the diffusive regime, in contrast, there are pre-localized 
states, which are weakly coupled to the leads. Due to their existence the distribution of 
the small resonance widths has a non-trivial behavior described by Eq.~(\ref{gammas}).  

\subsection{Wigner delay times distribution}

We turn now to the analysis of Wigner delay times and calculate their probability 
distribution. It can be shown that this distribution is related to the distribution
of reflection coefficients $R$ in the present of weak absorption. Absorption is one of 
the main ingredients in actual experimental situations and its theoretical understanding 
is of great importance. Unfortunately a comprehensive treatment of absorption is still
lacking. There are only very few reported analytical results  for 
the distribution of the reflection coefficient ${\cal P}(R)$ in the presence of absorption 
and all of them are either  within the regime of applicability of RMT \cite{KM00,BB01,SS03,DSF90,F03} 
or applied for quasi-1D geometry \cite{F03,MB96}. Here we will 
derive ${\cal P}(R)$ in the diffusive regime in two dimensions and in the weak absorption limit. 
In this limit it was shown \cite{BB01,DSF90} that the following relation between the proper 
Wigner delay times and reflection coefficients holds:
\begin{equation}
\label{rcoef}
R_q = 1-\tau_q/\tau_a
\end{equation}
where $\tau_q$ are the proper delay times (eigenvalues of the Wigner-Smith operator) and
$1/\tau_a$ is the absorption rate. Thus the knowledge of ${\cal P}(R)$ reduces to the
calculation of the distribution of proper Wigner delay times ${\cal P}(\tau_q)$ \cite{BB01}.

Bellow we make the standard assumption that the resulting distribution 
generated over different energies is equivalent with the one generated over different 
disorder realizations. Our starting point is the well known relation
\begin{equation}
\label{dtime}
\tau(\omega)=\sum_{n=1}^{L^2} \frac{\Gamma_n}{(\omega-\omega_n)^2 + \Gamma_n^2/4} 
\end{equation}
which connects the Wigner delay times and the poles of the $S-$matrix. 

\begin{figure}[!t]
\includegraphics[scale=0.5]{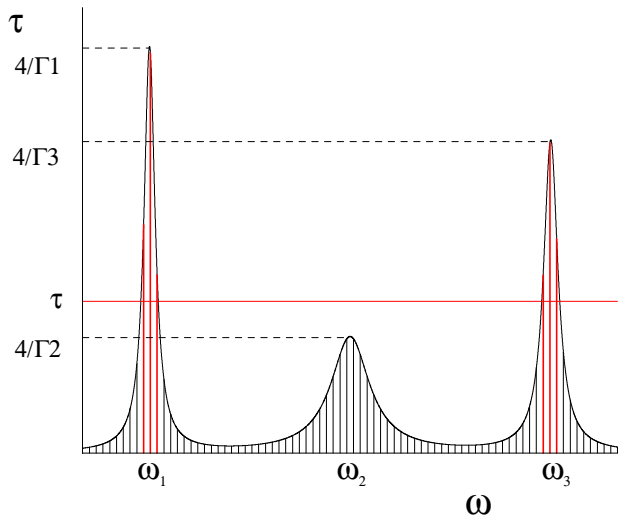}
\caption {\label{wign} 
Schematic plot for the Wigner delay time as a function of quasi-energy according to
Eq.~(\ref{dtime}).
}
\end{figure} 

\begin{figure}[!t]
\includegraphics[scale=0.5]{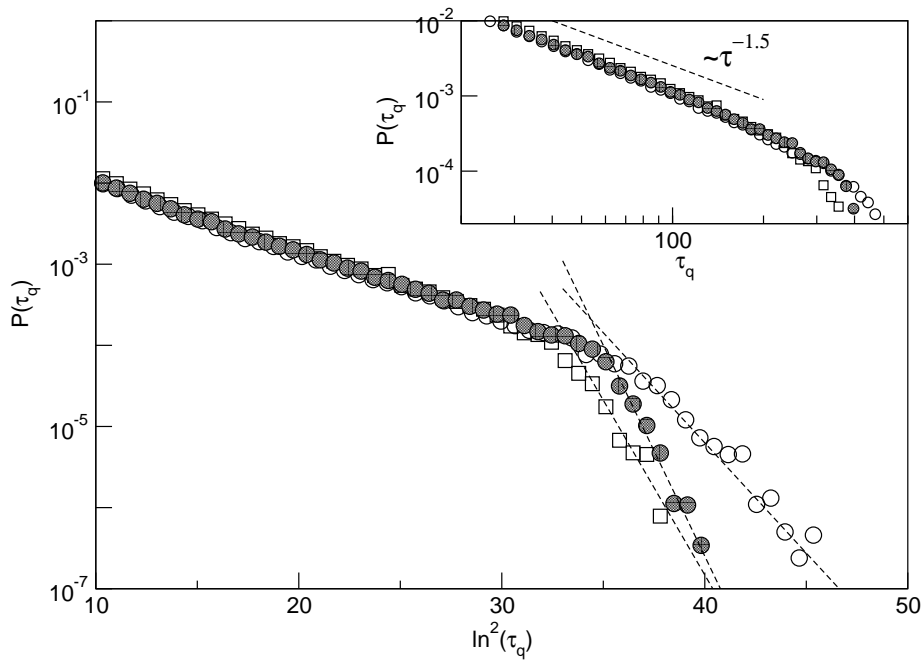}
\caption{\label{fig10}
The proper delay times distribution ${\cal P}(\tau_q)$ for $D=20.3$ ($\circ$) and 
$D=29.8$ ($\Box$ ). The ($\bullet$) correspond to $D=20.3$ but now with broken TRS. The 
dashed lines have slopes equal to $C_{\beta}$ extracted from the corresponding ${\cal P}
(\Gamma)$ (see Fig.~8). In the inset we report ${\cal P} (\tau_q )$ for moderate values 
of $\tau_q$ in a double logarithmic scale.}
\end{figure}

Let us start with the far tails. It is evident that large times $\tau(\omega)\sim 
\Gamma^{-1}_n$ corresponds to the cases when $\omega\simeq \omega_n$ and $\Gamma_n 
\ll 1$. In the neighborhood of these points, $\tau(\omega)$ can be approximated 
by a single Lorentzian (\ref{dtime}). Sampling the quasi-energies $\omega$ with 
step $\Delta \omega\ll \Gamma_{min}$ we calculate the number of points for which 
the time delay is larger than some fixed value $\tau$ (see Fig.~\ref{wign}). 
Assuming that the contribution of each Lorentzian is proportional to its width 
one can estimate this number as $\sum_{\Gamma_n < 1/\tau}\Gamma_n/\Delta E$. For 
the integrated distribution of delay times in the limit $\Delta \omega \rightarrow 0$
we obtain  
\begin{equation}
{\cal P}_{int} (\tau) \sim \int^{1/\tau} d\Gamma {\cal P}
(\Gamma)\Gamma 
\end{equation}
and by substituting the small resonance width asymptotic given by Eq.~(\ref{gammas}) 
we come out with the log-normal law of Eq.~(\ref{taus}) in agreement with our numerical 
findings reported in Fig.~\ref{fig10}.

Now we estimate the behavior of ${\cal P}(\tau)$ for $\tau \lesssim \Gamma_{cl}^{-1}$. 
In this regime many short-living resonances contribute to the sum (\ref{dtime}). We may
therefore consider $\tau$ as a sum of many independent positive random variables 
each of the type $\tau_n=\Gamma_n x_n$, where $x_n=\delta \omega_n^{-2}$. Assuming further
that $\delta \omega_n$ are uniformly distributed random numbers we find that the distribution
${\cal P}(x_n)$ has the asymptotic power law behavior $1/x_n^{3/2}$. As a next step we 
find that the distribution ${\cal P}(\tau_n)$ decays asymptotically as $1/\tau_n^{3/2}$ 
where we use that ${\cal P}(\Gamma_n)\sim 1/\Gamma_n^{3/2}$. Then the corresponding ${\cal P}
(\tau)$ is known to be a stable asymmetric Levy distribution $L_{\mu,1}(\tau)$ of 
index $\mu=1/2$ \cite{BG90} which has the form given in Eq.~(\ref{taus}) at the origin. 
We point out here that the asymptotic behavior ${\cal P}(\tau)\sim 1/\tau^{3/2}$ 
emerges also for chaotic/ballistic systems where the assumption of uniformly distributed 
$\delta \omega_n$ is the 
only crucial ingredient (see for example \cite{FyodSom97}). 

Since $\tau=\sum_{i=1}^M\tau_q$, we expect the behavior of the distribution of proper 
delay times ${\cal P}(\tau_q)$ to be similar to ${\cal P}(\tau)$ for large values of 
the arguments (for $\tau\gg1$ we have $\tau\sim \tau_q^{\rm max}$). Moreover, from 
the numerical point of view ${\cal P}(\tau_q)$ can be studied in a better way
because a larger set of data can be generated easily. Our numerical findings
for ${\cal P}(\tau_q)$ are reported in Fig.~\ref{fig10} and are in nice
agreement with Eq.~(\ref{taus}), even for moderate values of $\tau_q$. We stress here
that the dashed lines in Fig.~\ref{fig10}, have slopes equal to $C_{\beta}$
taken from the corresponding log-normal tails of ${\cal P}(\Gamma)$.

Finally we would like to compare our result (\ref{taus}) with the results known from the 
random matrix theory. Although the distributions of the proper delay times \cite{BFB97,SSS01}
and partial delay times  (defined as a derivative of the partial phase $\theta_i$ of the 
$S$-matrix with respect to energy) \cite{FyodSom97} have been calculated recently, there
is no explicit analytical formula for the distribution of the Wigner delay times for number
of open channels $M>2$ \cite{SFS01}. Nevertheless
using  consideration similar to one presented above (see the discussion for $\tau \lesssim 
\Gamma_{cl}^{-1}$) one can argue \cite{FyodSom97} that the part of ${\cal P}(\tau)$ for
 $\tau \lesssim \Gamma_{cl}^{-1}$ is the same in the RMT as stated in Eq.~(\ref{taus}).
However, the distribution of the large delay times in the RMT is expected to have a power 
law behavior ${\cal P}(\tau) \sim 1/\tau^{2+\beta M/2}$ with $M$ being the number 
of open channels. This is  in contrast with a log-normal tail stated in Eq.~(\ref{taus}). 

\section{Conclusions}

This paper is devoted to the investigation of chaotic and disordered systems characterized 
by the classical diffusion. Section V deals with study of the closed system. Specifically, 
we perform a detailed numerical analysis of the eigenfunction intensities ${\cal P}(y)$ of 
the two-dimensional kicked rotor on a torus. Our results indicated that the distribution 
${\cal P}(y)$ of generic quantum {\it dynamical} systems with diffusive classical limit is 
affected by the existence of {\it pre-localized} states. The deviations from RMT are well 
described by field theoretical methods developed for disordered systems. In particular, we
find  that the dependence of the tails of ${\cal P}_{\beta}(y)$ on TRS is described correctly
 by NLSM.

Section VI deals with the study of the open system. After introducing a scattering formalism
for the KR model we investigated the distribution of the resonance widths ${\cal P}(\Gamma)$ 
and Wigner delay times ${\cal P}(\tau)$. We obtain the forms of these distributions (log-
normal for large $\tau$ and small $\Gamma$, and power law in the opposite case) for different 
symmetry classes and show that they are determined by the underlying diffusive classical 
dynamics and by the existence of the pre-localized states. Our theoretical arguments are 
supported by extensive numerical calculations.

Although the arguments, we used to explain the behavior of ${\cal P}(\Gamma)$ and ${\cal P}(\tau)$, 
can be easy generalized to the three-dimensional case, the numerical test of these predictions has not
been still performed. Moreover the study of three-dimensional case would allow to investigate these 
distribution at the critical point of the metal-insulator transition. The first attempt to attack
this problem was done in Ref.\cite{KW02}, but a detailed  understanding is still required.

We thank I. E. Smolyarenko for useful comments. This work was supported by a grant from German-Israeli
 Foundation for Scientific Research and Development.
\section{Appendix I: The mean free path}

Let us consider the evolution operator of the two-dimensional kicked rotor
introduced in Eq.~(\ref{Uop}) (to simplify the calculations we put $T=1$)
\begin{equation}
\label{Uop-app}
U=e^{-iH_0({\cal L}_1,{\cal L}_2)/2}e^{-iV(\theta_1,\theta_2)}e^{-iH_0({\cal L}_1,{\cal L}_2)/2}.
\end{equation}
The set of the orthogonal eigenfunctions of the free Hamiltonian $H_0$ is
given by  plane waves in the angle representation
\begin{equation}
\phi_{(n_1,n_2)}(\theta_1,\theta_2)=\frac{1}{2\pi}e^{i(n_1 \theta_1 +n_2
  \theta_2)}.
\end{equation}
They are normalized in such a way that 
\begin{equation}
\int_0^{2\pi}\int_0^{2\pi}d\theta_1 d\theta_2
|\phi_{(n_1,n_2)}(\theta_1,\theta_2)|^2 =1.
\end{equation}
Let us denote by $U_{(k_1,k_2),(n_1,n_2)}$ the matrix elements of the evolution operator 
in this basis
\begin{equation}
U_{(k_1,k_2),(n_1,n_2)}=\<\phi_{(k_1,k_2)}|U|\phi_{(n_1,n_2)}\>.
\end{equation}
According to the definition of the evolution operator the modulus square of
its elements have the meaning of the probability to change the initial
momentum $(n_1,n_2)$ to the final momentum $(k_1,k_2)$ in one kick. Therefore one can 
define the {\it mean free path in momentum space}  \cite{DF88} by
\begin{equation}
\label{mfp-app} 
l_M^2=\sum_{r_1}\sum_{r_2}(r_1^2+r_2^2)|U_{(0,0),(r_1,r_2)}|^2.
\end{equation} 
Here we used the fact that $|U_{(k_1+m_1,k_2+m_2),(n_1+m_1,n_2+m_2)}|^2=
|U_{(k_1,k_2),(n_1,n_2)}|^2$, so without loss of generality one can take the
initial momentum equal to $(0,0)$.
In order to calculate the right hand side of Eq.~(\ref{mfp-app}) we first 
give an explicit expression to the matrix elements $U_{(k_1,k_2),(n_1,n_2)}$:
\begin{eqnarray}
\lefteqn{U_{(k_1,k_2),(n_1,n_2)}=}\nonumber\\
&&=\sum_{i,j,s,t} \<\phi_{(k_1,k_2)}|e^{-iH_0({\cal L}_1,{\cal  L}_2)/2}
|\phi_{(i,j)}\>\<\phi_{(i,j)}|e^{-iV(\theta_1,\theta_2)}|\phi_{(s,t)}\>\<\phi_{(s,t)}|e^{-iH_0({\cal L}_1,{\cal  L}_2)/2}|\phi_{(n_1,n_2)}\>=\nonumber\\
&&=e^{-i(H_0(k_1,k_2)+H_0(n_1,n_2))/2}\<\phi_{(k_1,k_2)}|e^{-iV(\theta_1,\theta_2)}|\phi_{(n_1,n_2)}\>
=\nonumber \\
&&= e^{-i(H_0(k_1,k_2)+H_0(n_1,n_2))/2}\frac{1}{4\pi^2}\int_0^{2\pi}\int_0^{2\pi}d\theta_1 d\theta_2
e^{-iV(\theta_1,\theta_2)}e^{i(n_1-k_1)\theta_1+i(n_2-k_2)\theta_2}
\end{eqnarray}
Taking $(k_1,k_2)=(0,0)$ and $(n_1,n_2)=(r_1,r_2)$ we obtain
\begin{equation}
U_{(0,0),(r_1,r_2)}= e^{-i(H_0(0,0)+H_0(r_1,r_2))/2}\frac{1}{4\pi^2}
\int_0^{2\pi}\int_0^{2\pi}d\theta_1 d\theta_2
e^{-iV(\theta_1,\theta_2)}e^{ir_1\theta_1}e^{ir_2\theta_2}
\end{equation}
The substitution of this expression into  Eq.~(\ref{mfp-app}) yields
\begin{equation}
\label{mfp-int}
 l_M^2=\sum_{r_1}\sum_{r_2}(r_1^2+r_2^2)\frac{1}{(4\pi^2)^2}
\int_0^{2\pi}\int_0^{2\pi} \int_0^{2\pi}\int_0^{2\pi}
d\theta_1 d\theta_2 d\tilde{\theta}_1 d\tilde{\theta}_2
e^{-i(V(\theta_1,\theta_2)-V(\tilde{\theta}_1,\tilde{\theta}_2))}
e^{ir_1(\theta_1-\tilde{\theta}_1)}e^{ir_2(\theta_2-\tilde{\theta}_2)}.
\end{equation} 
Taking into account that
\begin{equation}
r_1^2e^{ir_1(\theta_1-\tilde{\theta1}_1)}=\frac{\partial^2}{\partial \theta_1 
\partial \tilde{\theta}_1}e^{ir_1(\theta_1-\tilde{\theta1}_1)}
\end{equation}
and that the same is valid for $r_2$ the partial integration of the Eq.~(\ref{mfp-int})
gives
\begin{eqnarray}
l_M^2&=&\sum_{r_1}\sum_{r_2}\frac{1}{(4\pi^2)^2}
\int_0^{2\pi}\int_0^{2\pi} \int_0^{2\pi}\int_0^{2\pi}
d\theta_1 d\theta_2 d\tilde{\theta}_1 d\tilde{\theta}_2
\(\frac{\partial^2}{\partial \theta_1 \partial \tilde{\theta}_1}+
\frac{\partial^2}{\partial \theta_2 \partial \tilde{\theta}_2}\)\nonumber \\
&&e^{-i(V(\theta_1,\theta_2)-V(\tilde{\theta}_1,\tilde{\theta}_2))}
e^{ir_1(\theta_1-\tilde{\theta}_1)}e^{ir_2(\theta_2-\tilde{\theta}_2)}.
\end{eqnarray}
The summation of the exponents over $r_1$ and $r_2$ yields two
$\delta$-functions:
\begin{eqnarray}
l_M^2&=&\frac{1}{4\pi^2}
\int_0^{2\pi}\int_0^{2\pi} \int_0^{2\pi}\int_0^{2\pi}
d\theta_1 d\theta_2 d\tilde{\theta}_1 d\tilde{\theta}_2
\(\frac{\partial^2}{\partial \theta_1 \partial \tilde{\theta}_1}+
\frac{\partial^2}{\partial \theta_2 \partial \tilde{\theta}_2}\)\nonumber \\
&&e^{-i(V(\theta_1,\theta_2)-V(\tilde{\theta}_1,\tilde{\theta}_2))}
\delta(\theta_1-\tilde{\theta}_1)\delta(\theta_2-\tilde{\theta}_2)=\nonumber\\
&=&\frac{1}{4\pi^2}
\int_0^{2\pi}\int_0^{2\pi} \int_0^{2\pi}\int_0^{2\pi}
d\theta_1 d\theta_2 d\tilde{\theta}_1 d\tilde{\theta}_2
\(\frac{\partial V}{\partial \theta_1}(\theta_1,\theta_2)
\frac{\partial V}{\partial  \tilde{\theta}_1}(\tilde{\theta}_1,\tilde{\theta}_2)+
\frac{\partial V}{\partial \theta_2}(\theta_1,\theta_2)
\frac{\partial V}{\partial  \tilde{\theta}_2}(\tilde{\theta}_1,\tilde{\theta}_2)\)
\nonumber \\
&&e^{-i(V(\theta_1,\theta_2)-V(\tilde{\theta}_1,\tilde{\theta}_2))}
\delta(\theta_1-\tilde{\theta}_1)\delta(\theta_2-\tilde{\theta}_2)=\nonumber\\
&=&\frac{1}{4\pi^2}\int_0^{2\pi}\int_0^{2\pi} d\theta_1 d\theta_2
\[\(\frac{\partial V}{\partial \theta_1}(\theta_1,\theta_2)\)^2+
\(\frac{\partial V}{\partial \theta_2}(\theta_1,\theta_2)\)^2\]
\end{eqnarray}
The last expression can be written in a compact form
\begin{equation}
\label{mfp-nabla}
l_M^2=\frac{1}{4\pi^2}\int_0^{2\pi}\int_0^{2\pi} d\theta_1 d\theta_2
\|\vec{\nabla} V(\theta_1,\theta_2)\|^2
\end{equation}

Now we calculate the mean free path in the case where the potential 
$V(\theta_1,\theta_2)$ is given by  Eq.~(\ref{pot-2d}):
\begin{equation}
V(\theta_1,\theta_2) = k\(\cos(\theta_1) \cos(\theta_2) \cos({\alpha}) +
{1\over2} \sin(2\theta_1) \cos(2\theta_2) \sin({\alpha})\)
\end{equation}
Taking the derivative of this expression with respect to $\theta_1$
 and $\theta_2$ one has
\begin{eqnarray}
\(\frac{\partial V}{\partial \theta_1}(\theta_1,\theta_2)\)^2 &=&
k^2 ( \sin^2\theta_1 \cos^2\theta_2\cos^2\alpha+
 \cos^2 2\theta_1 \cos^2 2\theta_2\sin^2\alpha-\nonumber\\
&&-2 \sin\theta_1\cos 2\theta_1 \cos\theta_2\cos 2\theta_2\cos\alpha\sin\alpha)\nonumber \\
\(\frac{\partial V}{\partial \theta_2}(\theta_1,\theta_2)\)^2 &=&
k^2 (\cos^2\theta_1 \sin^2\theta_2\cos^2\alpha +
\sin^2 2\theta_1 \sin^2 2\theta_2\sin^2\alpha +\nonumber\\
&&+2\cos\theta_1\sin 2\theta_1 \sin\theta_2\sin 2\theta_2\cos\alpha\sin\alpha )
\end{eqnarray}
The integration over $\theta_1$ and $\theta_2$ yields for the mean free path
\begin{equation}
\label{mfp2-final}
l_M^2=\frac{1}{4\pi^2}k^2(\pi^2\cos^2\alpha+\pi^2\sin^2\alpha+
\pi^2\cos^2\alpha+\pi^2\sin^2\alpha)=\frac{k^2}{2}
\end{equation}
which is the same result with the one derived in \cite{DF88} for different potential 
$V(\theta_1,\theta_2)$.

\section{Appendix II: Diffusion coefficient in the random phase approximation}
\label{app-diff}

Here we give for completeness of the presentation the derivation of the diffusion coefficient for
our model (\ref{ham},\ref{pot-2d}) in the random phase approximation.
We start by writing the classical maps Eq.~(\ref{kr-maps2}) for the general
form of the potential $V(\theta_1,\theta_2)$:
\begin{eqnarray}
\thf &=& \theta_1(n)+\tau_1 T {\cal L}_1(n) \;\; \mbox{mod} \;\; 2 \pi \nonumber \\
\ths &=& \theta_2(n)+\tau_2 T {\cal L}_2(n) \;\; \mbox{mod} \;\; 2 \pi \nonumber \\
{\cal L}_1(n+1) &=& {\cal L}_1(n)-\frac{\partial V}{\partial \theta_1}(\thf,\ths)\nonumber\\
{\cal L}_2(n+1) &=& {\cal L}_2(n)-\frac{\partial V}{\partial \theta_2}(\thf,\ths)\nonumber\\
&& -\sin(2\thf)\sin(2\ths)\sin(\alpha)) 
\end{eqnarray}
The diffusion coefficient in momentum space is defined as
\begin{equation}
D= \lim_{t\rightarrow \infty}\frac{\<{\cal L}_1^2(t)+{\cal L}_2^2(t)\>}{t}
\end{equation}
The average in this expression is taken over an ensemble of trajectories, with
different initial conditions. Using the classical maps for ${\cal L}_1$ and
${\cal L}_2$ we obtain
\begin{eqnarray}
\label{diffcoeff-full-app}
D=\lim_{n\rightarrow \infty}\frac{1}{nT}\[\sum_{i=1}^{n}\<
\(\frac{\partial V}{\partial \theta_1}(\theta_1(i),\theta_2(i))\)^2+
\(\frac{\partial V}{\partial \theta_2}(\theta_1(i),\theta_2(i))\)^2\>+\right.\nonumber\\
+\left.\sum_{i=1}^n\sum^n_{j=1,j\neq i}\<
\frac{\partial V}{\partial \theta_1}(\theta_1(i),\theta_2(i)) 
 \frac{\partial V}{\partial \theta_1}(\theta_1(j),\theta_2(j))+
\frac{\partial V}{\partial \theta_2}(\theta_1(i),\theta_2(i)) 
 \frac{\partial V}{\partial \theta_2}(\theta_1(j),\theta_2(j))\>\]\nonumber\\
\end{eqnarray}
For the large values of the kicking strength $k$, in a good approximation one
can consider the phases $\theta_1(i)$ and $\theta_2(i)$ as random variables
which are uncorrelated for different $i$ and distributed uniformly in the
interval $[0,2\pi]$. Using this  random phase approximation it is easily to 
show that only the diagonal terms in Eq.~(\ref{diffcoeff-full-app}) give
non-zero contribution in the limit $n\rightarrow \infty$. Taking into account
that the distribution of the phases is uniform one can convert the sum for the 
diagonal terms into an integral. Finally we obtain
\begin{equation}
D=\frac{1}{4\pi^2T}\int_0^{2\pi}\int_0^{2\pi} d\theta_1 d\theta_2
\[\(\frac{\partial V}{\partial \theta_1}(\theta_1,\theta_2)\)^2+
\(\frac{\partial V}{\partial \theta_2}(\theta_1,\theta_2)\)^2\]
\end{equation}
Then the formula for the diffusion coefficient measured in  number of kicks
$D_k=TD$ is given by 
\begin{equation}
D_k=\frac{1}{4\pi^2}\int_0^{2\pi}\int_0^{2\pi} d\theta_1 d\theta_2
\|\vec{\nabla} V(\theta_1,\theta_2)\|^2
\end{equation}
which has exactly the same form as one appearing Eq.~(\ref{mfp-nabla}). Thus we obtain
that in the random phase approximation the following relation between the mean
free path and the diffusion coefficient is valid:
\begin{equation}
D_k=l_M^2=\frac{k^2}{2}
\end{equation}
Therefore changing the kicking strength we can easy tune the diffusion constant or
Thouless conductance (for disordered systems). This allows us to investigate
various regimes: ballistic, diffusive and localized.

\section{Appendix III: Unitarity of the $S$-matrix}

Let us rewrite the expression for $S$-matrix (\ref{smatrns}) in a more 
symmetric way. To this end we use a series expansion for  inverse operator 
in (\ref{smatrns}):
\begin{eqnarray}
\[U-\wexp-\wdw\]^{-1}=\[\(U-\wexp\)\(I-\frac{1}{U-\wexp}\wdw\)\]^{-1}=\nonumber\\
=\[I-\frac{1}{U-\wexp}\wdw\]^{-1}\frac{1}{U-\wexp}=
\sum_{k\ge 0}\(\frac{1}{U-\wexp}\wdw\)^k\frac{1}{U-\wexp}.
\end{eqnarray}
Substituting this expansion in Eq.~(\ref{smatrns}) we obtain
\begin{eqnarray}
S&=&\sum_{k\ge 0}WU\(\frac{1}{U-\wexp}\wdw\)^k\frac{1}{U-\wexp}W^{\dagger} =
\sum_{k\ge 0}\(WU\frac{1}{U-\wexp}W^{\dagger}\)^{k+1}=\nonumber \\
&=&\frac{WU\frac{1}{U-\wexp}W^{\dagger}}{I-WU\frac{1}{U-\wexp}W^{\dagger}}=
\frac{W\frac{U}{U-\wexp}W{\dagger}}{W\(I-\frac{U}{U-\wexp}\)W^{\dagger}}=
-\frac{W\frac{U}{U-\wexp}W^{\dagger}}{W\frac{\wexp}{U-\wexp}W^{\dagger}}
\end{eqnarray}
Now using the unitarity of the evolution operator $U$ we can calculate
the Hermitian conjugate $S$-matrix:
\newcommand{\pwexp}{e^{i\omega}}
\begin{eqnarray}
S^{\dagger}=-\frac{W\frac{U^{-1}}{U^{-1}-\pwexp}W^{\dagger}}{W\frac{\pwexp}
{U^{-1}-\pwexp}W^{\dagger}}=
-\frac{W\frac{\wexp}{\wexp-U}W^{\dagger}}{W\frac{U}{\wexp-U}W^{\dagger}}=
-\frac{W\frac{\wexp}{U-\wexp}W^{\dagger}}{W\frac{U}{U-\wexp}W^{\dagger}}=S^{-1}.
\end{eqnarray}
Thus the unitarity of the $S$-matrix is proven.

\end{document}